\documentclass[%
reprint,
superscriptaddress,
amsmath,amssymb,
aps,
prl,
]{revtex4-2}

\usepackage[dvipsnames]{xcolor}
\usepackage{graphicx}%
\usepackage[caption=false]{subfig}
\usepackage{wrapfig}
\usepackage{dcolumn}%
\usepackage{bm}%
\usepackage{hyperref}%
\usepackage{amsmath}
\usepackage{lineno}
\hypersetup{
  colorlinks,
  citecolor=blue,
  linkcolor=blue,
  urlcolor=blue}

\usepackage[per-mode=symbol]{siunitx}

\newcommand{\BR}[1]{}
\newcommand{\ak}[1]{}
\newcommand{\verify}[1]{}

\newcommand{\ks}{$\epsilon_\mathrm{KS},\ \gamma=0.1$\,eV}
\newcommand{\gw}{$E_{GW},\ \gamma=0.1$\,eV}
\newcommand{\gwD}{$E_{GW},\ \gamma^{GW}_{i,j,\mathbf{k}}$}

\begin{document}

\title{Capturing many-body effects in electrical conductivity of warm dense matter}

\author{Brian P. Robinson}
\affiliation{Department of Materials Science and Engineering, University of Illinois Urbana-Champaign, Urbana, IL, 61801, USA}
\affiliation{Center for Computing Research, Sandia National Laboratories, Albuquerque NM, USA}
\author{Alina Kononov}
\affiliation{Center for Computing Research, Sandia National Laboratories, Albuquerque NM, USA}
\author{Lucas J. Stanek}
\affiliation{Pulsed Power Sciences Center, Sandia National Laboratories, Albuquerque NM, USA}
\author{Andrew~D.~Baczewski}
\affiliation{Center for Computing Research, Sandia National Laboratories, Albuquerque NM, USA}
\author{Andr\'e Schleife}
\affiliation{Department of Materials Science and Engineering, University of Illinois Urbana-Champaign, Urbana, IL, 61801, USA}
\affiliation{Materials Research Laboratory, University of Illinois Urbana-Champaign, Urbana, IL, 61801, USA}
\affiliation{National Center for Supercomputing Applications, University of Illinois Urbana-Champaign, Urbana, IL, 61801, USA}
\author{Stephanie B. Hansen}
\affiliation{Pulsed Power Sciences Center, Sandia National Laboratories, Albuquerque NM, USA}

\begin{abstract}
Conductivity models for warm dense matter inform simulations of planetary structure and fusion experiments.
State-of-the-art conductivity calculations based on density functional theory approximate many-body physics and neglect electron-electron scattering lifetimes.
We introduce a many-body framework for electrical conductivity using the $GW$ approximation of the electronic self-energy.
For beryllium, improved transition energies yield a surprisingly large reduction in low-temperature DC conductivity, while electron-electron scattering primarily reduces high-temperature DC conductivity.
\end{abstract}

\maketitle

\textit{Introduction}---Accurate conductivity estimates for matter in extreme conditions are essential for modeling physical processes in inertial confinement fusion (ICF) and planetary dynamos~\cite{french:2012,nettelmann:2012,haines:2024}.
In ICF, DC conductivity governs the growth of yield-limiting Rayleigh-Taylor instabilities~\cite{lindl:1995,lindl:2004}, and
pulsed-power experiments~\cite{matzen:2005} relevant to magnetized liner inertial fusion (MagLIF) concepts require accurate DC conductivity models to infer current delivered.
The wide range of temperatures and densities accessed in high-energy-density (HED) experiments challenges models, particularly in the warm dense matter (WDM) regime, where thermal and quantum effects coexist and multiple scattering mechanisms contribute.

Two limiting theories commonly describe the DC conductivity of plasmas: the Ziman liquid-metal formalism~\cite{ziman:1961} in the low-temperature regime and the Spitzer-H\"{a}rm theory of classical plasmas~\cite{spitzer:1953} in the high-temperature regime~\cite{shaffer:2020}.
Ziman theory accounts for electron degeneracy and strong ion coupling, but typically neglects electron-electron scattering.
The Spitzer model assumes low degeneracy and weak ion coupling, but includes electron-electron scattering.
Parameterized analytical models like the Lee-More-Desjarlais (LMD) form~\cite{lee:1984,desjarlais:2017} can estimate conductivities through the crossover between the degenerate and classical limits.
These parameterized models~\cite{stanek:2024a} are often constrained by density functional theory (DFT)~\cite{mermin:1965} calculations, with multi-atom molecular dynamics (DFT-MD) simulations~\cite{desjarlais:2002,french:2022} serving as the 
state of the art for temperatures comparable to the Fermi energy and average-atom (DFT-AA) models~\cite{sterne2007equation,starrett2016kubo} efficiently 
extending DFT to high temperatures.

However, all DFT-based approaches approximate electron-electron interactions and DFT-AA further misses multi-center effects like lattice structure.
DFT-MD conductivity predictions do not approach the Spitzer limit at high temperatures~\cite{french:2022}, indicating inadequate treatment of electron-electron scattering.
The resulting DC conductivity errors could propagate into significant uncertainties in HED diagnostics ~\cite{stanek:2024a} and ICF target design.
While earlier work~\cite{reinholz:2015,ropke:2025} proposed capturing electron-electron scattering effects through a correction factor derived from approximate plasma models, a fully first-principles description remains absent.

This letter introduces a rigorous first-principles approach to improve the description of electron-electron interactions over standard DFT-MD conductivity calculations through the many-body self-energy within the $GW$ approximation \cite{hedin:1965}.
Starting from the DFT electronic structure, the self-energy provides many-body corrections to quasiparticle energies and the corresponding lifetimes.
While quasiparticle energy corrections are commonly considered in calculations of optical properties for ambient materials~\cite{shishkin:2007,reining2018gw} and excitation lifetimes enter spectral line broadening models for plasmas \cite{perrot1984hydrogen,gunter1991hydrogen}, their combined effect on WDM conductivity remains unexplored.
Our work demonstrates a framework for modifying the widely used Kubo-Greenwood (KG) formalism \cite{kubo:1957,greenwood:1958} for DFT-MD conductivity calculations to include quasiparticle effects.

As an exemplar, we consider the electrical conductivity of liquid beryllium near solid density at temperatures of $0.2$\,--\,\SI{6.8}{\electronvolt}.
We find systematic reductions in DC conductivity predictions, with quasiparticle energy corrections enhancing non-Drude behavior in the low-temperature limit and electron-electron scattering captured through quasiparticle lifetimes dominating at higher temperatures.
Therefore, standard DFT-based models---which neglect quasiparticle effects---overestimate DC conductivity values informing simulations of HED experiments.

\textit{Methods}---
The KG formalism~\cite{kubo:1957,greenwood:1958} determines electronic transport properties from the energies and occupations of electronic eigenstates along with the transition matrix elements among them.
In practice, the many-body wavefunctions and energies are typically replaced by single-particle Kohn-Sham (KS) states and eigenvalues computed from Mermin DFT~\cite{mermin:1965,calderin:2017}.
Within the KG formalism, the optical conductivity can be calculated from DFT results~\cite{holst:2011,calderin:2017} as
\begin{align}
  \sigma&(\omega)
  =\, \frac{2\pi e^2\hbar^3}{m_e^2 \Omega}\sum_\mathbf{k} w_\mathbf{k}\sum_{ij}\frac{f_{j,\mathbf{k}}-f_{i,\mathbf{k}}}{\epsilon_{i,\mathbf{k}}-\epsilon_{j,\mathbf{k}}} \notag \\
  &\times\langle\Psi_{i,\mathbf{k}}|\nabla|\Psi_{j,\mathbf{k}}\rangle\langle\Psi_{j,\mathbf{k}}|\nabla|\Psi_{i,\mathbf{k}}\rangle\delta(\epsilon_{i,\mathbf{k}}-\epsilon_{j,\mathbf{k}}-\hbar\omega),
  \label{eq:sigma}
\end{align}
where $\omega$ is the frequency, $e$ is the electron charge, $m_e$ is the electron mass, $\Omega$ is the simulation cell volume, and $w_\mathbf{k}$ is the quadrature weight corresponding to momentum $\mathbf{k}$.
The indices $i$ and $j$ correspond to band indices, while $f_{i,\mathbf{k}}$ is the Fermi occupation of the state, $\epsilon_{i,\mathbf{k}}$ is the eigenvalue of the state, and $\langle \Psi_{j,\mathbf{k}} | \nabla | \Psi_{i,\mathbf{k}} \rangle$ is the transition matrix element.
Finally, the $\delta$-function used for energy conservation can be approximated by a Lorentzian or Gaussian form with a broadening parameter $\gamma$.
Here, we use the Lorentzian function because it naturally represents the original many-body KG formalism \cite{calderin:2017}.

To re-introduce many-body physics missing from KS-DFT, we use the $GW$ approximation to compute the self-energy of electronic excitations as $\Sigma\approx iGW$, where $G$ is the single-particle Green's function and $W$ is the screened Coulomb interaction.
The real part of $\Sigma$ provides quasiparticle corrections to the KS energies, while the imaginary part describes finite quasiparticle lifetimes due to electron-electron interactions.
Thus, we obtain $GW$-corrected conductivities by replacing the KS-DFT eigenvalues $\epsilon_{i,\mathbf{k}}$ entering Eq.~\eqref{eq:sigma} with the $GW$-corrected energies $E^{GW}_{i,\mathbf{k}}$ and updating the corresponding occupations $f_{i,\mathbf{k}}$ to maintain Fermi statistics.
We also replace the \emph{ad hoc} Lorentzian broadening parameter with first-principles, transition-dependent broadening values from $GW$, $\gamma_{i,j,\mathbf{k}}^{GW}$, which arise from the excitation lifetimes of the initial and final states.

We applied our many-body framework to the case of liquid beryllium after equilibrating 54-atom simulation cells using DFT-MD.
To estimate DC conductivity, we extrapolated the dynamic conductivity to zero frequency by fitting to an even polynomial and then averaged results over multiple ionic configurations.
Section~\ref{sec:si:comp_details} of the SM describes more details about the computational approach.\@

\begin{figure}
  \includegraphics[width=\columnwidth]{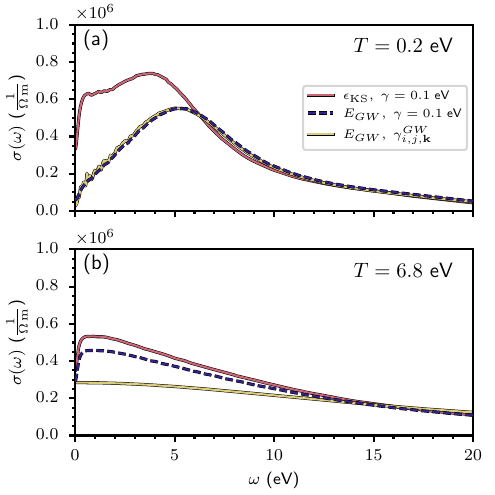}
  \caption{Optical conductivities predicted for \SI{1.97}{\gram\per\centi\meter\cubed} beryllium at temperatures of (a) 0.2 and (b) \SI{6.8}{\electronvolt} using the conventional KS-DFT treatment (\ks), after correcting KS energies with $GW$ (\gw), and after additionally including $GW$ broadening (\gwD).
  }\label{fig:therm_sigma}
\end{figure}

\textit{Results}---First, we investigate the influence of $GW$ corrections on the optical conductivity of warm-dense beryllium in Fig.~\ref{fig:therm_sigma}.
We focus on the near-solid density of \SI{1.97}{\gram\per\centi\meter\cubed} and two contrasting temperatures: \SI{0.2}{\electronvolt}, which lies slightly above the melting point, and \SI{6.8}{\electronvolt}, where we approach the feasibility limits of the costly $GW$ calculations.
For the lower temperature case shown in Fig.~\ref{fig:therm_sigma}(a), both of the $GW$-corrected curves fall below the KS conductivity for frequencies below \SI{5}{\electronvolt}.
Meanwhile, the $GW$ lifetimes do not appreciably alter the low-temperature results.
On the other hand, at the higher temperature considered in Fig.~\ref{fig:therm_sigma}(b), the $GW$ energy corrections lead to a modest shift in the conductivity at low frequencies but otherwise preserve the qualitative shape of the KS-DFT result.
However, including $GW$ lifetimes reduces the high-temperature dynamic conductivity for frequencies below \SI{10}{\electronvolt} and redistributes the spectral weight toward higher frequencies.
Overall, at low temperature, the set of eigenvalues and corresponding occupations affect the transport values most, while at high temperature, the choice of broadening method has a larger impact.

\begin{figure}
  \includegraphics[width=\columnwidth]{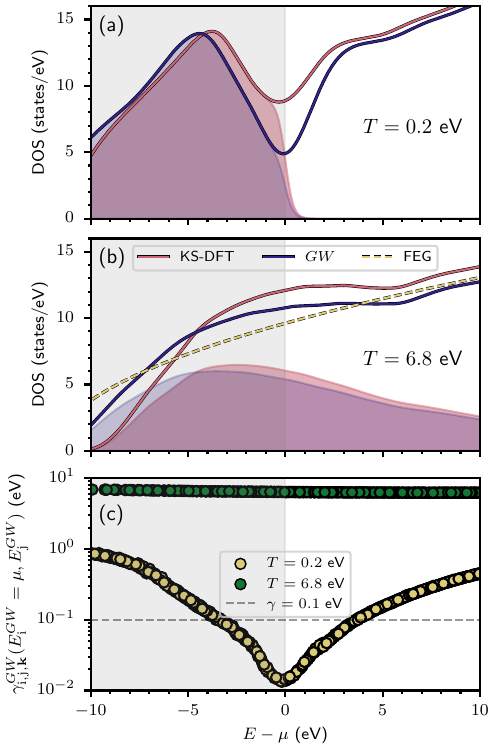}
  \caption{Electronic densities of states for \SI{1.97}{\gram\per\cubic\centi\meter} Be at temperatures of (a) $0.2$ and (b) \SI{6.8}{\electronvolt} predicted by KS-DFT (red), $GW$ (blue), and the free-electron gas model (yellow).
  The shaded regions show each DOS weighted by Fermi-Dirac occupations.
  In (c), the $GW$ broadening along $E^{GW}_i=\mu$ at both temperatures is compared to a typical static value of \SI{0.1}{\electronvolt} (dashed line).
  The $GW$ broadening for all energy transitions around the chemical potential are illustrated in Fig.~\ref{fig:therm_broad} of the SM.
  }\label{fig:therm_dos_broad}
\end{figure}

To understand the influence of $GW$ energy corrections on the low-temperature optical conductivity, we examine the electronic density of states (DOS) shown in Fig.~\ref{fig:therm_dos_broad}(a).
Both the KS-DFT and $GW$ DOS curves exhibit a significant dip near the chemical potential.
This well-known feature, the \textit{pseudogap}, is a prominent aspect of crystalline beryllium's electronic structure (see Fig.~\ref{fig:xc_dos} in the SM) and arises from strong $s-p$ band hybridization~\cite{haussermann:2001}.
Interestingly, the pseudogap persists above melt and leads to the non-Drude feature near $\omega=\SI{5}{\electronvolt}$ in the optical conductivity of Fig.~\ref{fig:therm_sigma}(a).
Isochorically heated beryllium features an even deeper pseudogap, and the resulting temperature-dependent non-Drude peak (see Sec.~\ref{sec:si:results:non_therm} of the SM) could be validated by e.g., low-angle scattering experiments at x-ray free electron laser facilities \cite{gawne:2024}.

In Fig.~\ref{fig:therm_dos_broad}(a), the KS-DFT pseudogap is shallower and narrower than the $GW$ pseudogap.
This $GW$ widening of the pseudogap is consistent with well-established behavior in ambient semiconductors, where $GW$ similarly increases the gap between valence and conduction bands, producing excellent agreement with experimental data~\cite{shishkin:2007}.
The larger pseudogap within $GW$ leads to fewer states near the chemical potential compared to KS-DFT.\@
Since states around the chemical potential dominate the optical conductivity at low temperatures, this result explains why the $GW$ optical conductivities in Fig.~\ref{fig:therm_sigma}(a) rapidly decrease for $\omega\lesssim \SI{5}{\electronvolt}$ compared to KS-DFT.

At temperatures well above melt, nearly all long-range order vanishes, thus also diminishing band structure effects like the $s-p$ hybridization.
The pseudogap is no longer present at $T=\SI{6.8}{\electronvolt}$ within either the KS-DFT or $GW$ DOS in Fig.~\ref{fig:therm_dos_broad}(b).
With the pseudogap gone, the KS-DFT and $GW$ curves predict a similar number of states around the chemical potential relative to the low-temperature case.
However, KS-DFT still predicts slightly more states near the chemical potential, which explains why the KS-DFT curve in Fig.~\ref{fig:therm_sigma}(b) predicts a slightly larger optical conductivity than $GW$.
Finally, at high temperatures the DOS begins to resemble the free-electron gas and follow an $E^{1/2}$ energy dependence within both levels of theory.
Since the DOS predicted by KS-DFT and $GW$ have the same general form at high temperature, the corresponding optical conductivities also have the same qualitative behavior.

While $GW$ energy corrections primarily impact the optical conductivity at low temperature, the $GW$ broadening plotted in Fig.~\ref{fig:therm_dos_broad}(c) primarily influences high-temperature conductivity.
At \mbox{$T=\SI{6.8}{\electronvolt}$}, the nearly constant $GW$ broadening of about \SI{6.5}{\electronvolt} significantly exceeds the values typically used in KS-DFT conductivity calculations (\SI{0.1}{\electronvolt} or less).
Other transition-energy pairs within $k_BT$ of the chemical potential show the same behavior (see Fig.~\ref{fig:therm_broad} in the SM).
This large broadening completely smears out any fine spectral features in Fig.~\ref{fig:therm_sigma}(b) and reduces the optical conductivity in the low-frequency limit while extending the high-frequency tail.
These effects reflect strong electron-electron scattering and short excitation lifetimes.

While the $GW$ broadening in Fig.~\ref{fig:therm_dos_broad}(c) shows little sensitivity to energy in the high-temperature case, the quadratic dependence on energy near the chemical potential at \mbox{$T=\SI{0.2}{\electronvolt}$} is consistent with Fermi liquid theory~\cite{vignale:2022,daligault:2017}.
The transition-dependent broadening from $GW$ falls below \SI{0.1}{\electronvolt} for transition energy pairs within \SI{4}{\electronvolt} of the chemical potential, with the $GW$ minimum about an order of magnitude below the static value used in the KS-DFT calculation.
Since Pauli blocking at low temperatures permits low-energy transitions only among states near the chemical potential, these small $GW$ broadening values lead to subtle noise in the low-frequency conductivity of Fig.~\ref{fig:therm_sigma}(a).
At higher frequencies, the $GW$ broadening becomes comparable to the constant \SI{0.1}{\electronvolt} value and thus does not noticeably modify the conductivity.

\begin{figure}
  \includegraphics[width=\columnwidth]{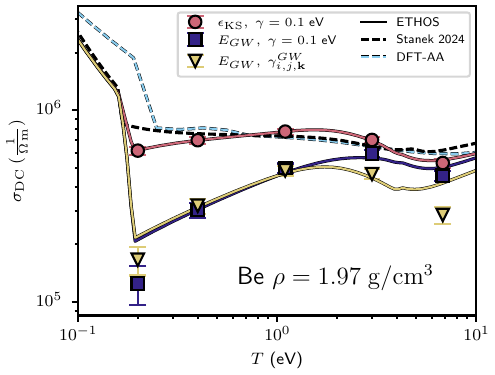}
  \caption{The temperature dependence of DC conductivity for Be at \SI{1.97}{\gram\per\cubic\centi\meter} computed using conventional KS-DFT (red circles), $GW$-corrected eigenvalues (blue squares), and $GW$-corrected dynamic broadening (yellow triangles).
  Solid curves correspond to the ETHOS~\cite{stanek:2024a} parameterization of the Lee-Moore-Desjarlais model~\cite{desjarlais:2001} constrained by our  KS-DFT (red), $GW$-corrected eigenvalues (blue), and $GW$-corrected dynamic broadening (yellow) data.
  The dashed black line is the interpolated and extrapolated ETHOS data by Stanek \mbox{\textit{et al.}}~\cite{stanek:2024a}, and the dashed light blue curve shows predictions from DFT-AA.
  Error bars estimate both the statistical and systematic errors in the predictions as described in Sec.~\ref{sec:si:convergence} of the SM.\@
  }\label{fig:therm_dc}
\end{figure}

The zero-frequency limit of the optical conductivity determines DC conductivity, an essential transport coefficient for HED simulations.
The $GW$-corrected DC conductivities in Fig.~\ref{fig:therm_dc} show the same trends discussed previously:
at low temperatures, $GW$ energy corrections are more significant than $GW$ broadening effects, while high temperatures show the opposite behavior.
Overall, $GW$ corrections reduce DC conductivity predictions for Be by a factor of 2\,--\,5 relative to KS-DFT results.
We find that this effect also holds at other densities of $1.70$--\SI{2.10}{\gram\per\centi\meter\cubed} for $T=\SI{0.40}{\electronvolt}$ (see Fig.~\ref{fig:therm_dc_rho} in the SM).

In Fig.~\ref{fig:therm_dc}, we compare our DC conductivity results to several computationally efficient models capable of tabulating values over the wide range of conditions needed for HED simulations.
First, we compare to our ETHOS~\cite{stanek:2024a} parameterization of the LMD model~\cite{desjarlais:2001} constrained by a multi-fidelity dataset including experimental data in the solid regime, KS-DFT or $GW$-corrected data in the liquid regime, and DFT-AA data in the plasma regime.
For each KS-DFT or $GW$-corrected dataset, we take the median parameters from a ten-dimensional posterior distribution obtained through Bayesian inference.
The ETHOS parameterizations do not fully capture the low- and high-temperature limits of our $GW$-corrected data, motivating future improvements to our fit model.

In Fig.~\ref{fig:therm_dc}, we also compare our DC conductivity results to two preexisting computationally efficient models.
DFT-AA computes the conductivity from the electronic structure of a single representative Be atom (see Sec.~\ref{sec:si:aa} of the SM for more details), with known shortcomings at low temperature.
The Stanek \textit{et al.} curve is an interpolated ETHOS parameterization constrained by DFT-AA and previous KS-DFT calculations at nearby densities~\cite{stanek:2024a}. 
As expected, our KS-DFT results agree fairly well with both DFT-AA and Stanek \textit{et al.}~\cite{stanek:2024a}, with slightly smaller conductivities near the melt transition and slightly larger curvature over the given temperature range.  
The curvature is particularly important because it influences the evolution of electrothermal instabilities~\cite{peterson:2012,yu:2023}.
The $GW$-corrected values are systematically lower than predicted by the other models and have significantly more curvature in this range.

\textit{Conclusions}---We find that many-body effects beyond DFT systematically lower DC electrical conductivity predictions for beryllium near solid density by factors of \mbox{2\,--\,5}.
At low temperatures, $GW$ corrections to the single-particle KS energies widen the pseudogap, leading to fewer available transitions near the chemical potential.
At high temperatures, short excitation lifetimes lead to a large spectral broadening that redistributes dynamic conductivity away from the DC limit.

Since electron-electron scattering rates generally increase with temperature \cite{daligault:2017}, we anticipate similar high-temperature 
corrections
for other materials.
In particular, our approach could help resolve recent 
questions
about incorrect limiting behavior within KS-DFT at high temperatures~\cite{french:2022}.
In contrast, the conductivity reductions predicted at low temperatures result from peculiarities of beryllium's electronic structure
and would benefit from experimental investigations~\cite{ofori-okai:2025,cochrane:2016}.

For ideal metals, we expect closer agreement between KS-DFT and $GW$-corrected low-temperature conductivities.
We also expect our approach to extend to $d$-band metals like copper \cite{clerouin:2005}, where single-shot $G_0W_0$ predictions for ambient conditions show better agreement with experiment by capturing $d-d$ intraband decay channels missed in simpler treatments~\cite{marini:2002}.
Within our $GW$ conductivity formalism, the resulting increase in broadening would redistribute spectral weight into the Lorentzian tails, which could either increase or decrease the DC conductivity depending on the nature of any non-Drude behavior.

Although our $GW$ approach includes additional physics beyond DFT and has been extensively validated for optical properties of ambient semiconductors~\cite{shishkin:2007}, some aspects may require further scrutiny in the warm-dense regime.
First, self-consistently solving for $G$ and $W$ beyond single-shot calculations could influence our conductivity predictions.
So-called vertex corrections accounting for higher-order many-body interactions could also have an impact.
Earlier work on the homogeneous electron gas~\cite{takada:2001} and solids~\cite{marini:2002} suggests that the effects of self-consistency and vertex corrections largely cancel near the chemical potential, leaving low-temperature electrical conductivities approximately unchanged.
However, the behavior of these effects at higher temperatures, where energies away from the chemical potential increasingly participate in transport, remains unexplored.
These effects may also significantly impact low-temperature thermal conductivity estimates, as discussed in Sec.~\ref{sec:si:results:therm} of the SM.

\begin{acknowledgments}
\textit{Acknowledgments}---We thank %
Alex White,
Charlie Starrett,
Cody Melton,
Jackson White,
Nathaniel Shaffer,
Normand Modine,
and
Suxing Hu
for numerous helpful and thought-provoking discussions.
We also acknowledge support from Sandia National Laboratories' Laboratory Directed Research and Development (LDRD) Projects No.\ 218456 and 233196 and the US Department of Energy Science Campaign 1.
This work was performed, in part, at the Center for Integrated Nanotechnologies, an Office of Science User Facility operated for the U.S.\ Department of Energy (DOE) Office of Science.

This article has been co-authored by employees of National Technology \& Engineering Solutions of Sandia, LLC under Contract No. DE-NA0003525 with the U.S. Department of Energy (DOE). The authors own all right, title and interest in and to the article and are solely responsible for its contents. The United States Government retains and the publisher, by accepting the article for publication, acknowledges that the United States Government retains a non-exclusive, paid-up, irrevocable, world-wide license to publish or reproduce the published form of this article or allow others to do so, for United States Government purposes. The DOE will provide public access to these results of federally sponsored research in accordance with the DOE Public Access Plan \url{https://www.energy.gov/downloads/doe-public-access-plan}.
\end{acknowledgments}

\bibliography{ref}

\clearpage
\renewcommand\thesection{S\arabic{section}}
\renewcommand{\theHsection}{S\thesection}
\renewcommand\thefigure{S\arabic{figure}}
\renewcommand{\theHfigure}{S\thefigure}
\renewcommand\theequation{S\arabic{equation}}
\renewcommand{\theHequation}{S\theequation}
\renewcommand\thetable{S\arabic{table}}
\renewcommand{\theHtable}{S\thetable}
\setcounter{figure}{0}
\setcounter{equation}{0}
\setcounter{section}{0}
\setcounter{table}{0}
\setcounter{secnumdepth}{5}
\onecolumngrid

\begin{center}
  \large\textbf{Supplemental Material: Capturing many-body effects in electrical conductivity of warm dense matter}\\[6pt]
  B. P. Robinson$^{1,2}$,
  A. Kononov$^{2}$,
  L. J. Stanek$^{5}$,
  A. D. Baczewski$^{2}$,
  A. Schleife$^{1,3,4}$,
  S. B. Hansen$^{5}$\\[4pt]
  \small
  $^{1}$Department of Materials Science and Engineering, University of Illinois at Urbana-Champaign, Urbana, IL 61801, USA\\
  $^{2}$Center for Computing Research, Sandia National Laboratories, Albuquerque, NM 87123, USA\\
  $^{3}$Materials Research Laboratory, University of Illinois at Urbana-Champaign, Urbana, IL 61801, USA\\
  $^{4}$National Center for Supercomputing Applications, University of Illinois at Urbana-Champaign, Urbana, IL 61801, USA\\
  $^{5}$Pulsed Power Sciences Center, Sandia National Laboratories, Albuquerque, NM 87123, USA
\end{center}
\vspace{1.5em}

\section{Methodological Details}
\label{sec:si:comp_details}

\subsection{DFT-MD and conventional Kubo-Greenwood formalism}
\label{sec:si:dft}

\begin{wrapfigure}{r}{0.45\textwidth}
    \includegraphics[width=0.45\columnwidth]{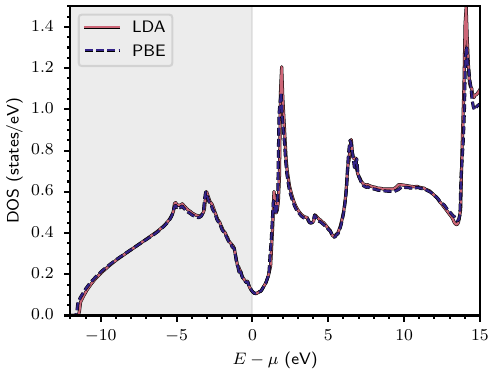}
    \caption{\label{fig:xc_dos}
    The density of states for \SI{1.97}{\gram\per\cubic\centi\meter} crystalline beryllium (HCP lattice) at an electronic temperature of~\SI{0.5}{\electronvolt} as predicted using KS-DFT with the LDA and PBE XC functionals.
    } 
\end{wrapfigure}

Our finite-temperature KS-DFT study of beryllium used Mermin DFT~\cite{mermin:1965} within \textsc{vasp}~\cite{kresse:1996,kresse:1996a,kresse:1999}.
Depending on the specific calculation, the electronic exchange and correlation (XC) in the DFT Hamiltonian was governed by either the local-density approximation (LDA)~\cite{kohn:1965} or the generalized-gradient approximation (GGA) parameterized by Perdew, Burke, and Ernzerhof (PBE)~\cite{perdew:1996}.
Since we found only minimal differences between LDA and PBE predictions for the electronic density of states within \SI{10}{\electronvolt} of the chemical potential (see Fig.~\ref{fig:xc_dos}), we do not expect significant sensitivities to the choice of XC functional.
The electron-ion interactions were described by the projector-augmented wave method (PAW)~\cite{blochl:1994}.
To ensure an accurate description of the core states, we used a four-electron PAW potential, which treats both the $1s$ and $2s$ electrons as valence electrons.

While the main text focuses on thermalized conditions where the ions and electrons are in thermal equilibrium at the same temperature, in Sec.~\ref{sec:si:results:non_therm} we also consider isochorically heated conditions where the electronic temperature is finite ($T_e>0$) but the ions occupy crystalline lattice sites corresponding to zero ionic temperature ($T_i=0$).
For the isochorically heated case, we used the LDA XC functional for all calculations.
The $T_e$ values ranged from $0.1$ to \SI{7}{\electronvolt} and determined KS occupations through the Fermi-Dirac distribution.
We used the two-atom hexagonal close-packed (HCP) primitive cell for the crystalline ionic geometry, where the ionic coordinates were fully relaxed. 
Fitting the total energy as a function of lattice parameters to the Murnaghan equation of state~\cite{murnaghan:1937} gave equilibrium lattice parameters of $a=b=$ 2.23 \AA\ and $c=$ 3.53 \AA\, corresponding to a density of \SI{1.97}{\g\per\cubic\centi\meter}.
We performed subsequent conductivity calculations, including for the thermalized cases, at this density.

In the thermalized case, we considered five temperatures above melt: $T_i=T_e=0.2$, $0.4$, $1.1$, $3.0$, and \SI{6.8}{\electronvolt}.
To equilibrate the ionic system, we performed KS-DFT molecular dynamics (DFT-MD) simulations with a 54-atom unit cell, a $\Gamma$-centered $\mathbf{k}$-point grid, and a plane-wave energy cutoff of \SI{600}{\electronvolt}.
For each temperature, we started in the canonical (NVT) ensemble where the velocity rescaling thermostat was used for several picoseconds (ps) with the LDA XC functional.
Then, we continued the DFT-MD calculations in the microcanonical (NVE) ensemble using PBE, where we ran for another several ps, ensuring that the energy and temperature at each time-step were not drifting significantly away from their equilibrium values.
Finally, we determined adequate time-step strides for extracting uncorrelated snapshots by finding where the velocity autocorrelation function crosses zero.
We report these values for each temperature in Table~\ref{tab:md_data}.

\begin{table}[!h]
    \begin{center}
    \caption{\label{tab:md_data}
    Properties of the DFT-MD calculations at each temperature. 
    NVT and NVE times indicate the length of the DFT-MD calculations in each ensemble.
    Minimum stride is the time at which the velocity autocorrelation function crosses zero, and snapshots refers to the number of uncorrelated configurations used for the conductivity calculations. 
    }
    \begin{tabular}{c c c c c} 
        \hline
        $T$ (eV) 
        & \; NVT Time (ps)
        & \; NVE Time (ps)
        & \; Minimum Stride (fs)
        & \; Snapshots \\ [0.5ex] 
        \hline
        0.2 & 6  & 9  & 16  & 8  \\ 
        0.4 & 6  & 7  & 17  & 9  \\ 
        1.1 & 2  & 2  & 79  & 11 \\ 
        3.0 & 2  & 4  & 100 & 11 \\ 
        6.8 & 2  & 2  & 172 & 5  \\ 
        \hline
    \end{tabular}
    \end{center}
\end{table}

We computed electronic transport properties with the Kubo-Greenwood (KG) formalism~\cite{kubo:1957,greenwood:1958},
where the many-body wavefunctions are typically replaced by single-particle states from DFT~\cite{calderin:2017}.
The optical and thermal conductivities can be found within the KG formalism through the frequency-dependent Onsager coefficients~\cite{holst:2011,melton:2024} of the form
\begin{align}
    L_{mn}^{\alpha,\beta}(\omega) 
    =\, &{(-1)}^{m+n} \frac{2\pi e^{3-m} \hbar^3}{(e T_e)^{n-1} m_e^2 V} \notag %
    \sum_{i,j,\mathbf{k}} w_\mathbf{k} 
    \left( \frac{\epsilon_{i,\mathbf{k}}+\epsilon_{j,\mathbf{k}}}{2} - \mu \right)^{m+n-2} \notag \\
    &\times \Re \left[ \Lambda_{i,j,\mathbf{k}}^{\alpha}\Lambda_{j,i,\mathbf{k}}^{\beta} \right] %
    \times \frac{\Delta f_{j,i,\mathbf{k}}}{\Delta\epsilon_{i,j,\mathbf{k}}}
    \, \delta(\Delta\epsilon_{i,j,\mathbf{k}} - \hbar \omega),
    \label{eq:onsager_SI}
\end{align}
where $L_{mn}^{\alpha,\beta}$ is the Onsager coefficient specified by the coefficient indices $m$ and $n$ and Cartesian indices $\alpha$ and $\beta$.
Further, $\omega$ is the frequency, $e$ is the electron charge, $T_e$ is the electronic temperature, $m_e$ is the electron mass, and $V$ is the simulation cell volume.
Each term in the sum accounts for the contribution from a transition between states $|\Psi_{i,\mathbf{k}}\rangle$ and $|\Psi_{j,\mathbf{k}}\rangle$, where $i$ and $j$ correspond to band indices and $\mathbf{k}$ is a reciprocal-space point within the Brillouin zone with quadrature weight $w_\mathbf{k}$. 
Thus, $\Delta f_{j,i,\mathbf{k}}=f_{j,\mathbf{k}}-f_{i,\mathbf{k}}$ is the difference of the occupations, $\Delta\epsilon_{i,j,\mathbf{k}} = \epsilon_{i,\mathbf{k}}-\epsilon_{j,\mathbf{k}}$ is the difference of the eigenvalues, 
and $\Lambda_{i,j,\mathbf{k}}^{\alpha}$ is the matrix element for the two states.
The matrix elements are taken in the longitudinal form---since, for the nonlocal pseudopotentials used in this work, only the longitudinal expression is formally correct\cite{melton:2024}---and are written as
\begin{equation}
  \Lambda_{i,j,\mathbf{k}}^{\alpha}
  =
  \Big\langle\Psi_{i,\mathbf{k}}\Big|\partial_\alpha-\frac{m_e}{\hbar^2}[\hat{V},\hat{r}_\alpha]\Big|\Psi_{j,\mathbf{k}}\Big\rangle
  \label{eq:matrix_elements},
\end{equation}
where %
$\partial_\alpha$ is the partial derivative with respect to Cartesian index $\alpha$, $\hat{V}$ is the potential energy operator, and $\hat{r}_\alpha$ is the $\alpha$ component of the position operator.
Finally, the $\delta$-function used for energy conservation can be approximated by a Lorentzian or Gaussian.
In this work, we use the Lorentzian function since it is consistent with lifetime broadening effects and is more representative of the original many-body Kubo-Greenwood formulation~\cite{calderin:2017}.
That is, we use
\begin{equation}
  \delta(\Delta\epsilon_{i,j,\mathbf{k}} - \hbar \omega) \approx \frac{1}{\pi}\frac{\gamma}{{(\Delta\epsilon_{i,j,\mathbf{k}}-\hbar\omega)}^2+{\gamma}^2},
\label{eq:simple_lorentzian}
\end{equation}
where $\gamma=\SI{0.1}{\electronvolt}$ is an ad hoc broadening parameter.

The real part of the electrical and thermal conductivities can be found from the Onsager coefficients through
\begin{gather}
  \sigma(\omega) = \frac{1}{3} \sum_{\alpha} L_{11}^{\alpha\alpha}(\omega), \label{eq:re_sigma} \\
  \kappa = \frac{1}{3} \lim_{\omega \to 0} \sum_{\alpha} \left[ L_{22}^{\alpha\alpha}(\omega)
  - \frac{L_{12}^{\alpha\alpha}(\omega) L_{21}^{\alpha\alpha}(\omega)}{L_{11}^{\alpha\alpha}(\omega)} \right]. \label{eq:kappa}
\end{gather}
To predict the DC conductivity, we must fit the frequency-dependent conductivity and extrapolate to zero frequency.
This curve fitting approach is necessary because finite-size effects inevitably cause an artificial drop in the conductivity at low frequencies \cite{french:2022,melton:2024}. %
Since Be does not follow the ideal Drude form, we fit to an even polynomial~\cite{melton:2024}.
More specifically, we tested both a fourth and sixth degree polynomial of the form 
\begin{equation}
    \sigma(\omega)\approx a_0+a_2\,\omega^2+a_4\,\omega^4\;(+a_6\,\omega^6).
    \label{eq:sigma_fit}
\end{equation}
We performed the fits for different frequency ranges where the lower bound was between $0.1$ and \SI{0.5}{\electronvolt} and the upper bound was between $2.0$ and \SI{5.0}{\electronvolt}.
Then, we selected the fit with the largest $R^2$ value to estimate the DC conductivity.

\subsection{$GW$ and modified Kubo-Greenwood formalism}
\label{sec:si:gw}

To capture the influence of excitations on the electronic structure, 
we went beyond DFT by employing the $GW$ approximation~\cite{hedin:1965}, which captures the many-body electron interactions through the self-energy ($\Sigma$).
In the $GW$ approximation, the self-energy is approximated as $\Sigma\approx iGW$, where $G$ is the single-particle Green's function and $W$ is the screened Coulomb interaction.
The real part of the self-energy relates to the quasiparticle energy shift, while the imaginary part describes the lifetime of the state due to electron-electron (e-e) interactions.
We used the low-scaling $GW$ algorithm~\cite{rojas:1995,liu:2016,kaltak:2020} implemented in \textsc{vasp}, which uses the Matsubara formalism~\cite{matsubara:1955} to perform $GW$ calculations at finite temperatures.
The DFT eigenvalues and wavefunctions computed with LDA served as the starting point for the $GW$ calculations. %
The DC conductivity predicted using a PBE starting point agrees exceptionally well with the LDA-based workflow (see Section~\ref{sec:si:convergence}), indicating that our results are not sensitive to the XC functional used in the initial DFT step.
Due to computational limitations, we ran single-shot ($G_0W_0$) calculations, where the self-energy was computed from a single $GW$ iteration.

To go beyond the commonly used single-particle description of Section~\ref{sec:si:dft} and include many-body effects from the $GW$ approximation, we first replaced the DFT eigenvalues and occupations in Eq.~\eqref{eq:onsager_SI} with corrected energies $E^\mathrm{GW}_{i,\mathbf{k}}$ and the corresponding Fermi occupations.
$E^\mathrm{GW}_{i,\mathbf{k}}$ is the real part of the complex quasiparticle energies $E_{i,\mathbf{k}}^\mathrm{QP}$, which can be estimated either by root searching or by linearizing the self-energy around the DFT single-particle eigenvalues.
Root searching involves finding the roots of the equation
\begin{equation}
  \left\langle\psi_{i,\mathbf{k}}\left|T+V_\mathrm{ext}+V_\mathrm{H}+\Sigma(\omega)\right|\psi_{i,\mathbf{k}}\right\rangle - \omega
  =
  0,
  \label{eq:qp_zeros}
\end{equation}
where $|\psi_{i,\mathbf{k}}\rangle$ is the fixed DFT starting-point wavefunction and the solution to Eq.~\eqref{eq:qp_zeros} is the complex quasiparticle energy.
Alternatively, the quasiparticle energies can be efficiently estimated by Taylor expanding the self-energy around the DFT energies~\cite{liu:2016}:
\begin{equation}
  E_{i,\mathbf{k}}^\mathrm{QP}
  \approx
  \epsilon_{i,\mathbf{k}}
  +
  Z_{i,\mathbf{k}}
  \langle\psi_{i,\mathbf{k}}|\Sigma(\epsilon_{i,\mathbf{k}})-V_{xc}|\psi_{i,\mathbf{k}}\rangle,
  \label{eq:qp_linear}
\end{equation}
where the renormalization factor is
\begin{equation}
  Z_{i,\mathbf{k}}
  \approx
  \left(
  1 - \frac{\partial \operatorname{Re}\Sigma(\omega)}{\partial\omega}\bigg|_{\omega=\epsilon_{i,\mathbf{k}}}
  \right)^{-1}.
  \label{eq:qp_Z_linear}
\end{equation}
In this work, we found no significant difference between the quasiparticle energies obtained from the two methods, so we proceeded with the more computationally efficient linearized method. %

While the real part of the complex quasiparticle energy gives the $GW$-corrected state energy, 
\begin{equation}
    E_{i,\mathbf{k}}^\mathrm{GW} = \mathrm{Re}[E_{i,\mathbf{k}}^\mathrm{QP}],
\end{equation}
the imaginary part determines the quasiparticle lifetime broadening
\begin{equation}
  \gamma_{i,\mathbf{k}}^{GW}
  \equiv
  \frac{\hbar}{2\tau_{i,\mathbf{k}}}
  =
  -\operatorname{Im} E^{\mathrm{QP}}_{i,\mathbf{k}}
  \label{eq:gw_broad}
\end{equation}
where $\gamma_{i,\mathbf{k}}^{GW}$ is the half-width at half-maximum of the quasiparticle peak in the single-particle spectral function
and $\tau_{i,\mathbf{k}}$ is the corresponding lifetime.
Within the linearized quasiparticle approximation of Eq.~\eqref{eq:qp_linear}, $\operatorname{Im} E^{\mathrm{QP}}_{i,\mathbf{k}}$ is estimated from the renormalized imaginary part of the self-energy, $Z_{i,\mathbf{k}} \operatorname{Im}\Sigma(\epsilon_{i,\mathbf{k}})$.

Typical KG calculations use a constant broadening value between $0.01$ and \SI{1.00}{\electronvolt} within the Lorentzian of Eq.~\eqref{eq:simple_lorentzian}.
However, in this work we introduce the $GW$ transition-dependent broadening, which is found by summing the broadening of the initial and final states,
\begin{equation}
  \gamma_{i,j,\mathbf{k}}^{GW}=\gamma_{i,\mathbf{k}}^{GW}+\gamma_{j,\mathbf{k}}^{GW}.
  \label{eq:gamma_gw}
\end{equation}
This transition-dependent broadening enters the KG formula by replacing the delta function in Eq.~\eqref{eq:onsager_SI} with the Lorentzian function described by
\begin{equation}
  F \left(\Delta E^{GW}_{i,j,\mathbf{k}}-\hbar\omega,\gamma^{GW}_{i,j,\mathbf{k}}\right) 
  = \frac{1}{\pi}\frac{\gamma^{GW}_{i,j,\mathbf{k}}}{{\left(\Delta E^{GW}_{i,j,\mathbf{k}}-\hbar\omega\right)}^2+{\left(\gamma^{GW}_{i,j,\mathbf{k}}\right)}^2}.
  \label{eq:gw_lorentzian}
\end{equation}
Thus, the GW-informed Onsager coefficients are given by
\begin{align}
    L_{mn}^{\alpha,\beta}(\omega) 
    =\, &{(-1)}^{m+n} \frac{2\pi e^{3-m} \hbar^3}{(e T_e)^{n-1} m_e^2 V} \notag %
    \sum_{i,j,\mathbf{k}} w_\mathbf{k} 
    \left( \frac{E^{\mathrm{GW}}_{i,\mathbf{k}}+E^{\mathrm{GW}}_{j,\mathbf{k}}}{2} - \mu^{\mathrm{GW}} \right)^{m+n-2} \notag \\
    &\times \Re \left[ \Lambda_{i,j,\mathbf{k}}^{\alpha}\Lambda_{j,i,\mathbf{k}}^{\beta} \right] %
    \times \frac{\Delta f^{\mathrm{GW}}_{j,i,\mathbf{k}}}{\Delta E^{\mathrm{GW}}_{i,j,\mathbf{k}}}
    \, F \left(\Delta E^{GW}_{i,j,\mathbf{k}}-\hbar\omega,\gamma^{GW}_{i,j,\mathbf{k}}\right) .
    \label{eq:onsager_gw}
\end{align}

Finally, since the non-thermalized case involves a simple two-atom primitive cell, the conductivity calculation requires a dense $50\times 50\times 50$ Monkhorst-Pack~\cite{monkhorst:1976} $\mathbf{k}$-mesh to converge (see Sec.~\ref{sec:si:convergence}).
This $\mathbf{k}$-mesh is much denser than what is feasible for a $GW$ calculation, where we used a $14\times 14\times 14$ $\Gamma$-centered grid.
This grid was the densest $\mathbf{k}$-point grid that was computationally feasible for our $GW$ calculations.
As a result, the number of states sampled in the $GW$ calculation is less than those sampled in the conductivity calculation, and thus we cannot replace the KS-DFT values with the corresponding $GW$ values directly.
To address this, we performed a simple linear interpolation of the $GW$ data as a function of the KS-DFT eigenvalues and used these interpolated $GW$ values in our conductivity study.
This interpolation procedure was not necessary for the thermalized case considered in the main text, where the $GW$ and conductivity calculations used the same $\mathbf{k}$-mesh.

\subsection{KS-DFT and $GW$ convergence tests and uncertainty estimation}
\label{sec:si:convergence}

To ensure accurate conductivity predictions and estimate uncertainties in the conductivities of thermalized and isochorically heated Be, we performed extensive sensitivity testing with respect to the exchange-correlation (XC) potential starting point, pseudopotential, reciprocal-space grid, number of atoms in the simulation cell, plane-wave energy cutoff, number of bands, and number of atomic configurations averaged.
The results of these tests are summarized in Table~\ref{tab:therm_dc_convergence} for the thermalized case and Tables~\ref{tab:primitive_gw_convergence} and~\ref{tab:nkpts_convergence} for isochorically heated Be.

\begin{table}[!h]
    \begin{center}
    \caption{\label{tab:therm_dc_convergence}
    Conductivity convergence tests for thermalized beryllium at a density of \SI{1.97}{\g\per\cubic\centi\meter}.
    The sensitivity of $\sigma_\mathrm{DC}$ to the number of ions, the plane-wave cutoff energy, reciprocal-space sampling, and the number of electronic bands used in the conductivity calculation is shown for both conventional KS-DFT calculations (see Sec.~\ref{sec:si:dft}) and GW-informed calculations (see Sec.~\ref{sec:si:gw}).
    The reciprocal-space grids are specified in terms of the total number of $\mathbf{k}$ points ($3^4$ or $4^3$), $\Gamma$-centered or Monkhorst-Pack (MP) grid, and the number of symmetrically-irreducible $\mathbf{k}$ points in parentheses.
    }
    \renewcommand{\arraystretch}{1.15}
    \begin{tabular}{c c c c c c c c} 
        \hline
        $T$ (eV) 
        & \; Eigenvalues
        & \; Broadening (eV)
        & \; Ions
        & \; Energy cutoff (eV)
        & \; $\mathbf{k}$ mesh
        & \; Bands
        & \; $\sigma_\mathrm{DC}\;(10^{5}\,(\Omega\,m)^{-1})$ \\ [0.5ex] 
        \hline
        3.0 & KS-DFT  & 0.1                            & 54  & 600 & $3^3$ $\Gamma$ (14)  & 378  & 7.02 \\ 
        3.0 & KS-DFT  & 0.1                            & 128 & 900 & $4^3$ MP (32)        & 931  & 7.30 \\ 
        \hline
        0.2 & $GW$    & $\gamma_{GW}^{i,j,\mathbf{k}}$ & 54  & 600 & $4^3$ $\Gamma$ (36)  & 224  & 1.69 \\ 
        0.2 & $GW$    & $\gamma_{GW}^{i,j,\mathbf{k}}$ & 54  & 600 & $4^3$ $\Gamma$ (36)  & 896  & 1.71 \\ 
        \hline
        6.8 & $GW$    & $\gamma_{GW}^{i,j,\mathbf{k}}$ & 54  & 600 & $3^3$ $\Gamma$ (14)  & 504  & 2.82 \\ 
        6.8 & $GW$    & $\gamma_{GW}^{i,j,\mathbf{k}}$ & 54  & 600 & $3^3$ $\Gamma$ (14)  & 2268 & 2.86 \\ 
        \hline
    \end{tabular}
    \end{center}
\end{table}

In Table~\ref{tab:therm_dc_convergence}, we report the main DC conductivity convergence tests we performed.
At $T=\SI{3.0}\electronvolt$, we estimated the KS-DFT DC conductivity with a set of parameters that are feasible to extend to the $GW$ approach and a more stringent set of parameters beyond $GW$ feasibility limits.
We find an absolute error of $2.8\times10^4\;{(\Omega m)}^{-1}$ and a $3.8\%$ relative error between the two calculations, which we consider acceptable given the large increase in computational expense for the more converged calculation.
To be conservative in our uncertainty estimates, we included this absolute difference within the conductivity error bars for all the conditions we studied.
At $T=0.20$ and \SI{6.8}{\electronvolt}, we find that including 224 and 504 bands, respectively, in the conductivity calculations yields well-converged $GW$-informed DC conductivities.
We again conservatively estimate uncertainties by including the larger of the two absolute differences, $4.0\times10^3\;{(\Omega m)}^{-1}$, in the error bars for all our conductivity results.

In addition, we averaged the dynamic conductivity over snapshots and computed the frequency-dependent standard deviation across the conductivity snapshots to estimate the uncertainty from thermal fluctuations.
Furthermore, we obtained the uncertainty in the zero-frequency extrapolation (Eq.~\eqref{eq:sigma_fit}) from the fit covariance matrix.
\ak{Brian, do you agree with the revised phrasing?}
\BR{Close, but I did not include the freq-dependent SD in my DC conductivity uncertainty.}
We computed a single uncertainty estimate for each DC conductivity value by taking the root-sum-square of the fit uncertainty and the convergence errors from Table~\ref{tab:therm_dc_convergence}.
These uncertainties are shown as the error bars in Fig.~\ref{fig:therm_dc} of the main text and in supplementary conductivity results in Sec.~\ref{sec:si:results:therm}.

\begin{figure}[h!]
  \includegraphics[width=0.75\columnwidth]{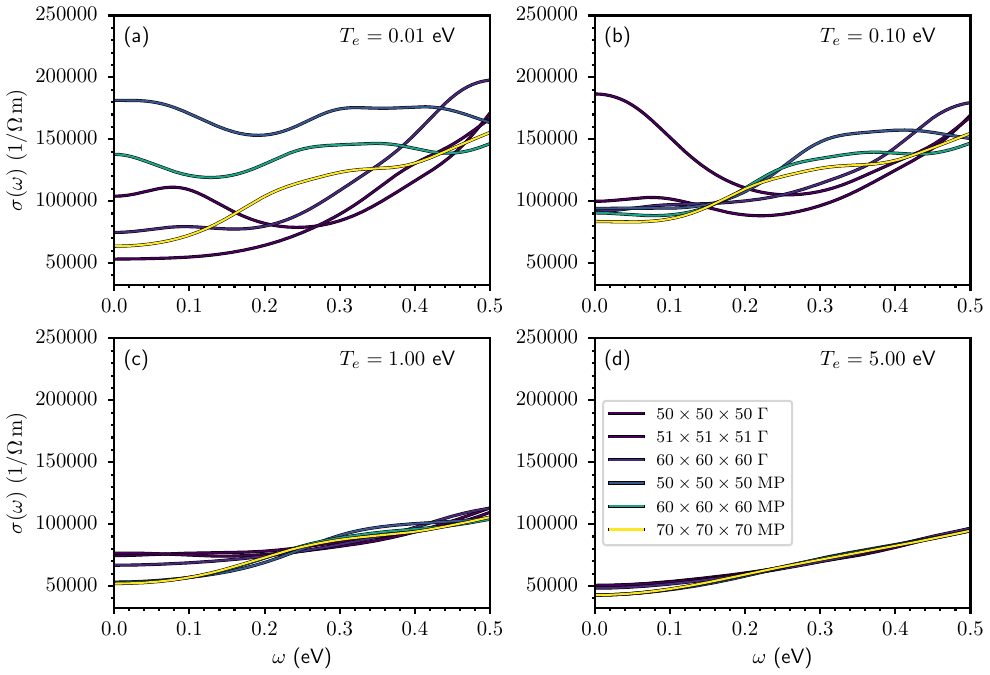}
  \caption{Low-frequency conductivity predictions for crystalline beryllium at \SI{1.97}{\gram\per\centi\meter\cubed} using KS-DFT with different $\Gamma$-centered and Monkhorst-Pack (MP) $\mathbf{k}$-point grids.
  Four different electronic temperatures between $0.01$ and \SI{5}{\electronvolt} are compared.
}\label{fig:nontherm_dc_kpoint}
\end{figure}

\begin{table}[!h]
    \centering
    \caption{\label{tab:primitive_gw_convergence}
    $GW$ convergence and sensitivity tests for the DC conductivity of isochorically heated beryllium at \SI{1.97}{\g\per\cubic\centi\meter} and $T_e=\SI{1.00}{\electronvolt}$.
    Relative differences between $\sigma_\mathrm{DC}$ predictions are listed as each parameter is varied while the others remain fixed.
    }
    \renewcommand{\arraystretch}{1.15}
    \begin{tabular}{c | c c | c}
        \hline
        Varied parameter & Test 1 & Test 2 & $\sigma_\mathrm{DC}$ relative difference (\%) \\ [0.5ex]
        \hline
        XC starting point & PBE & LDA & 0.37 \\
        Pseudopotential & sv & sv\_GW & 0.09 \\
        Energy Cutoff (eV) & 700 & 800 & 0.02 \\
        $GW$ Bands & 256 & 384 & 0.17 \\
        \hline
    \end{tabular}
\end{table}

For isochorically heated beryllium, we find in Table~\ref{tab:primitive_gw_convergence} that the $GW$-informed conductivity does not strongly depend on pseudopotential hardness or the choice of LDA or PBE XC functional as the $GW$ calculation's starting point.
We also find that using 192 total $\mathbf{k}$-points, an energy cutoff of \SI{700}{\electronvolt}, and 256 total bands for the 2-atom primitive unit cell all lead to well-converged results.
Finally, as illustrated in Fig.~\ref{fig:nontherm_dc_kpoint} and Table~\ref{tab:nkpts_convergence}, the DC conductivity for crystalline Be at low electronic temperatures is very difficult to converge with respect to the $\mathbf{k}$-point density.
We see this most clearly in the $T_e=\SI{0.01}{\electronvolt}$ case of Fig.~\ref{fig:nontherm_dc_kpoint}(a), where even with $44100$ symmetry-irreducible $\mathbf{k}$-points (corresponding to a $70^3$ grid), the zero-frequency limit is still not converged.
Although this behavior improves with increased temperature (see Fig.~\ref{fig:nontherm_dc_kpoint}(b)-(d)), very dense $\mathbf{k}$-point grids are required to fully converge the DC limit, increasing computational costs.
This difficulty arises because of sharp features in cold beryllium's Fermi surface, which contribute significantly to low-temperature transport but are challenging to sample adequately.

\begin{table}[!h]
    \begin{center}
    \caption{\label{tab:nkpts_convergence}
    DC conductivity estimates for isochorically heated beryllium at \SI{1.97}{\gram\per\cubic\centi\meter} and different electronic temperatures as predicted by KS-DFT using different size $\Gamma$-centered and Monkhorst-Pack~\cite{monkhorst:1976} (MP) $\mathbf{k}$-point grids.
    The number of symmetrically-irreducible $\mathbf{k}$-points is listed in parentheses for each case.
    }
    \renewcommand{\arraystretch}{1.15}
    \begin{tabular}{c c c c c c c} 
        \hline
        & \multicolumn{6}{c}{$\sigma_\mathrm{DC}\;(10^{5}\,(\Omega\,m)^{-1})$} \\
        \cline{2-7}
        \rule{0pt}{2.6ex}$T_e$ (eV)\vphantom{$^{3}$} 
        & \; $50^{3}\,\Gamma$ (6084)
        & \; $51^{3}\,\Gamma$ (6318) 
        & \; $60^{3}\,\Gamma$ (10261)
        & \; $50^{3}$ MP (16250)
        & \; $60^{3}$ MP (27900)
        & \; $70^{3}$ MP (44100) \\ [0.5ex] 
        \hline
        0.01 & 1.04 & 0.53 & 0.75 & 1.81 & 1.38 & 0.64 \\ 
        0.10 & 1.00 & 1.87 & 0.92 & 0.94 & 0.91 & 0.84 \\ 
        1.00 & 0.75 & 0.77 & 0.67 & 0.53 & 0.53 & 0.52 \\ 
        5.00 & 0.50 & 0.51 & 0.48 & 0.43 & 0.43 & 0.43 \\ 
        \hline
    \end{tabular}
    \end{center}
\end{table}

Overall, these careful convergence tests ensure an acceptable balance of accuracy and computational efficiency were maintained.
We report the final set of parameters used for thermalized and isochorically heated calculations in Tables~\ref{tab:therm_calc_table} and~\ref{tab:non_therm_calc_table}, respectively.
For all conditions, we chose sufficiently dense imaginary frequency and time grids for the $GW$ calculations to ensure the error in the electron number was below $10^{-5}$.

\begin{table}[h!]
\centering
\caption{\label{tab:therm_calc_table}
Converged parameter sets used in conductivity calculations for thermalized beryllium with a $\rho=$\SI{1.97}{\g\per\cubic\centi\meter} mass density.
The number of ions in the simulation cell, the plane-wave kinetic energy cutoff, the total number of bands included in the $GW$ and KS-DFT calculations, the number of imaginary frequency and time grid points $N_\omega$, and the $\mathbf{k}$-point mesh are listed for each temperature.
}
\renewcommand{\arraystretch}{1.15}
  \begin{tabular}{c c c c c c c}
    \hline
    $T$ (eV) 
    &\; Ions 
    &\; Energy cutoff (eV) 
    &\; $GW$ Bands 
    &\; KG Bands 
    &\; $N_\omega$ 
    &\; $\mathbf{k}$ mesh \\
    \hline
    0.2 & 54 & 600 & 1080 & 224 & 12 & $4^3$ $\Gamma$ (36) \\
    0.4 & 54 & 600 & 1080 & 232 & 12 & $4^3$ $\Gamma$ (36) \\
    1.1 & 54 & 600 & 2310 & 260 & 12 & $3^3$ $\Gamma$ (14) \\ 
    3.0 & 54 & 600 & 4352 & 378 & 12 & $3^3$ $\Gamma$ (14) \\
    6.8 & 54 & 600 & 6528 & 504 & 12 & $3^3$ $\Gamma$ (14) \\
  \hline
  \end{tabular}
\end{table}

\begin{table}[h!]
\centering
\caption{\label{tab:non_therm_calc_table}
Converged parameter sets used in conductivity calculations for isochorically heated beryllium with a $\rho=$\SI{1.97}{\g\per\cubic\centi\meter} mass density.
The number of ions in the simulation cell, the plane-wave kinetic energy cutoff, the total number of bands included in the $GW$ and KS-DFT calculations, the number of imaginary frequency and time grid points $N_\omega$, the density of the $\mathbf{k}$-point mesh and total $\mathbf{k}$-points in the $GW$ calculation, and the $\mathbf{k}$-point meshes used in the $GW$ and KG conductivity calculations are listed for each temperature.
}
\renewcommand{\arraystretch}{1.15}
  \begin{tabular}{c c c c c c c c}
    \hline
    $T$ (eV) 
    &\; Ions 
    &\; Energy cutoff (eV) 
    &\; $GW$ Bands 
    &\; KG Bands 
    &\; $N_\omega$ 
    &\; $GW$ $\mathbf{k}$ mesh 
    &\; KG $\mathbf{k}$ mesh \\
    \hline
    0.10 & 2 & 700 & 256 & 12 & 20 & $14^3$ $\Gamma$ (192) & $50^3$ MP (16250) \\
    0.20 & 2 & 700 & 256 & 12 & 20 & $14^3$ $\Gamma$ (192) & $50^3$ MP (16250) \\
    0.75 & 2 & 700 & 256 & 12 & 20 & $14^3$ $\Gamma$ (192) & $50^3$ MP (16250) \\
    1.00 & 2 & 700 & 256 & 12 & 20 & $14^3$ $\Gamma$ (192) & $50^3$ MP (16250) \\
    3.00 & 2 & 700 & 256 & 16 & 12 & $14^3$ $\Gamma$ (192) & $50^3$ MP (16250) \\
    7.00 & 2 & 700 & 256 & 24 & 12 & $14^3$ $\Gamma$ (192) & $50^3$ MP (16250) \\
  \hline
  \end{tabular}
\end{table}

\subsection{Average-atom calculations}
\label{sec:si:aa}
The average-atom model used in this work generates Kohn-Sham orbitals for all electrons of a single atom in a self-consistent, spherically symmetric potential~\cite{Liberman1979,Wilson2006}. It uses the LDA exchange-correlation functional prescribed in~\cite{Hedin_1971} and a static structure factor based on the quantum Ornstein-Zernike equation to describe ion correlations~\cite{starrett2014hedp}. The DC electrical conductivity is computed using the Ziman-Evans approach \cite{ziman:1961,sterne2007equation,hansen:2022} with a T-matrix differential cross section to account for strong collisions. Electron-electron collisions are included based on the approximation given in ~\cite{Potekhin_ee}.
While the AA model provides reasonable estimates for conductivities in warm and hot dense matter~\cite{stanek:2024}, it does not accurately account for multi-center effects such as band structure and lattice structure at temperatures near and below melt.

\section{Supplementary Results}
\subsection{Thermalized beryllium}
\label{sec:si:results:therm}

\begin{figure}[h!]
  \includegraphics[width=\columnwidth]{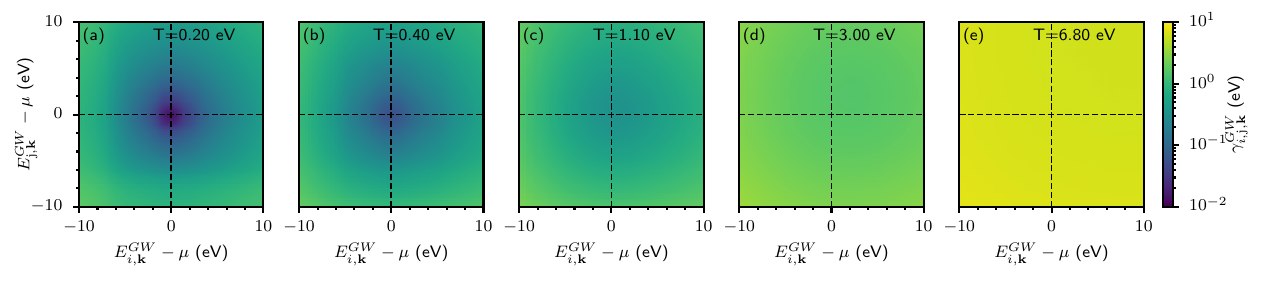}
  \caption{The temperature-dependent $GW$-informed Lorentzian broadening of thermalized \SI{1.97}{\g\per\cubic\centi\meter} beryllium as function of the initial ($E_{i,\mathbf{k}}$) and final ($E_{j,\mathbf{k}}$) energies in the optical transition.
  These values were computed according to Eqs.~\eqref{eq:gw_broad} and \eqref{eq:gamma_gw}, and they enter $GW$-informed conductivities through the Lorentzian broadening function of Eq.~\eqref{eq:gw_lorentzian}.
  }\label{fig:therm_broad}
\end{figure}

First, Fig.~\ref{fig:therm_broad} shows the full set of $GW$ broadening results entering the $GW$-informed conductivity predictions.
The data shown in Fig.~\ref{fig:therm_dos_broad}(c) of the main text represents line cuts from Fig.~\ref{fig:therm_broad} along $E_{i,\mathbf{k}}=\mu$.
At each temperature, the behavior for $E_{i,\mathbf{k}}$ within $k_B T$ of the chemical potential is very similar to the $E_{i,\mathbf{k}}=\mu$ case.
In particular, at high temperatures, the $GW$-informed broadening is nearly independent of the state energies.
As mentioned in the main text, this behavior matches what we expect from Fermi liquid theory~\cite{vignale:2022,daligault:2017}, where the broadening has approximately an energy-squared and temperature-squared dependence shown by
\begin{equation}
  \gamma^\mathrm{FL}
  \propto
  \frac{{\left(E-E_F\right)}^2+{\left(\pi k_B T\right)}^2}{1+e^{-\beta(E-E_F)}},
  \label{eq:fl_eq}
\end{equation}
where $E_F$ is the Fermi energy and is inverse temperature defined by $\beta=1/k_B T$.
\ak{Including the equation from Fermi liquid theory would help clarify the expected trends. As written, it sounds like the broadening goes as $E^2T^2$, which is not consistent with the data in Fig.~\ref{fig:therm_broad}.}
\BR{Good point - added}
The energy-squared dependence is illustrated clearly in Fig.~\ref{fig:therm_broad}(a) and (b), where the broadening increases quadratically with energy relative to the chemical potential.
We see the temperature dependence by noticing the broadening values in Fig.~\ref{fig:therm_broad}(a) change more with respect to energy compared to the relatively constant broadening in Fig.~\ref{fig:therm_broad}(e).

\begin{figure}
  \includegraphics[width=0.5\columnwidth]{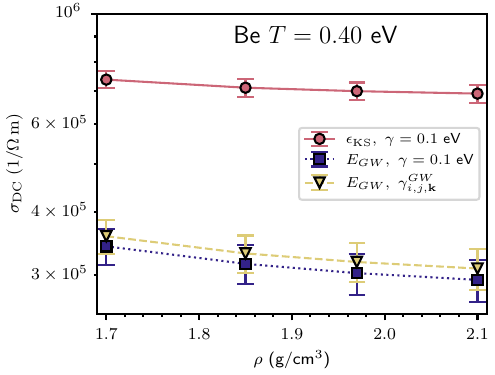}
  \caption{The mass-density dependence of DC conductivity for Be at $T=\SI{0.4}{\electronvolt}$ computed using conventional KS-DFT (\ks), $GW$-corrected eigenvalues (\gw), and $GW$-informed dynamic broadening (\gwD).
  Error bars estimate both the statistical and systematic errors in the predictions as described in Sec.~\ref{sec:si:convergence}.
  }\label{fig:therm_dc_rho}
\end{figure}

In Fig.~\ref{fig:therm_dc_rho}, we show how our DC conductivity predictions at $T=\SI{0.4}{\electronvolt}$ change as a function of mass density between 1.70 and \SI{2.10}{\g\per\cubic\centi\meter}.
All methods predict a modest decrease in DC conductivity with increasing density.
The $GW$-based predictions appear slightly more sensitive to density, likely arising from more pronounced pseudogap deepening and widening with increasing density for $GW$ relative to KS-DFT.
\BR{Yes I did}
However, this trend is quite weak relative to the uncertainties in the conductivity estimates.

\begin{figure}
  \includegraphics[width=\columnwidth]{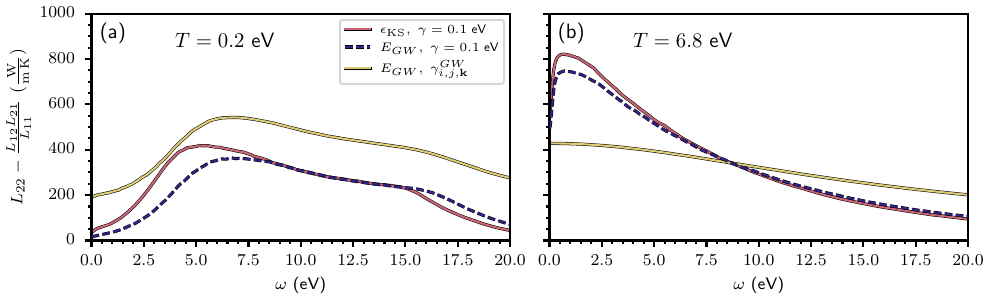}
  \caption{The frequency-dependent thermal conductivities predicted for \SI{1.97}{\g\per\cubic\centi\meter} beryllium at (a) $T=0.2$ and (b) \SI{6.8}{\electronvolt} using the standard KS-DFT approach (\ks), correcting KS energies with $GW$ (\gw), and after additionally including $GW$-informed broadening (\gwD).
  }\label{fig:therm_kappa}
\end{figure}

In the main text we only discussed electrical conductivity results, but we also computed thermal conductivity from the Onsager coefficients as given by Eq.~\eqref{eq:kappa}.
We show frequency-dependent values in Fig.~\ref{fig:therm_kappa}, which we extrapolated to zero frequency to predict the thermal conductivities.
At $T=\SI{0.2}{\electronvolt}$, including $GW$-informed broadening leads to a thermal conductivity prediction that is approximately $4$ and $8$ times larger than the KS-DFT and $GW$-energy corrected results that use a constant \SI{0.1}{\electronvolt} broadening (see Fig.~\ref{fig:therm_kappa}(a)).
This large thermal conductivity estimate originates from the ${[(\epsilon_{i,\mathbf{k}}+\epsilon_{j,\mathbf{k}})/2-\mu]}^{m+n-2}$ factor in Eq.~\eqref{eq:onsager_SI} when $m=n=2$.
This ``energy squared'' factor amplifies the $GW$-broadened Lorentzian tails for large-energy transitions, where the broadening values are between $1$\,--\,\SI{10}{\electronvolt} (see Fig.~\ref{fig:therm_broad}).
We do not see this behavior for the KS-DFT or $GW$ curves using a constant \SI{0.1}{\electronvolt} broadening because the significantly smaller broadening generates a much smaller Lorentzian tail that suppresses contributions from large-energy transitions.

Lorentzian tail effects from $GW$ broadening in the zero-frequency limit is not seen at high temperatures (see Fig.~\ref{fig:therm_kappa}(b)) because the amplification of the tail from the energy-squared term is much smaller than the low-energy transition values.
We show the nonphysical nature of the low-temperature result by calculating the Lorenz number
\begin{equation}
  L=\frac{e^2}{k_B^2}\frac{\kappa}{\sigma T}
  \label{eq:lorenz_number}
\end{equation}
using the Wiedemann-Franz law, which states that the ratio of thermal conductivity to DC conductivity is proportional to temperature~\cite{franz:1853}.
In the degenerate limit, the Sommerfeld expansion of the Fermi integrals finds $L_0=\pi^2/3$~\cite{ashcroft:1976,french:2022}.
Our Lorenz number to $L_0$ ratio are reported in Table~\ref{tab:L_L0_table}, where the $T=\SI{0.20}{\electronvolt}$ fully-$GW$ informed value is over $20$ times larger $L_0$.
This result shows that the amplification of the high-energy Lorentzian tails from the $GW$-broadening is nonphysical.

\begin{table}[htbp]
\centering
\caption{Lorenz number ratio $L/L_0$ as a function of temperature and method.}
\label{tab:L_L0_table}
\renewcommand{\arraystretch}{1.15}
\setlength{\tabcolsep}{8pt}
\begin{tabular}{c c c c}
\hline
$T$ (eV) & KS-DFT & GW & GW+$\gamma_{i,j,\mathbf{k}}^{GW}$ \\
\hline
0.2 & 1.56 & 3.47 & 21.48 \\
0.4 & 1.20 & 1.72 & 4.30 \\
3.0 & 0.91 & 0.99 & 0.99 \\
6.8 & 0.80 & 0.85 & 0.78 \\
\hline
\end{tabular}
\end{table}

We speculate that the nonphysical estimate of the fully-$GW$ thermal conductivity is due to a combination of inaccurate broadening predictions far from the chemical potential and missing physics.
X-ray photoelectron spectroscopy (XPS) of beryllium near ambient conditions measured a $1s$ energy between $-110$ and \SI{-112}{\electronvolt} below the chemical potential and estimated Lorentzian half width at half maximum (HWHM) between $0.1$\,--\,\SI{0.6}{\electronvolt}~\cite{mallinson:2013,pereyaslavtsev:2025}.
Our $GW$ calculation at a similar condition predicted a $1s$ energy of \SI{-114}{\electronvolt} in good agreement with XPS, but a HWHM between $2$\,--\,\SI{3}{\electronvolt}.
This HWHM enters the Onsager coefficient calculations as the state-dependent broadening in Eq.~\eqref{eq:gamma_gw}.
As previously discussed, high-energy transitions are not significant for DC conductivity, but are important in thermal conductivity due to the energy-squared term.
The order-of-magnitude overprediction of the $1s$ HWHM by $GW$ contributes to the overprediction of the $GW$ thermal conductivity due to Lorentzian tail amplification.
We believe this overprediction of the HWHM is not exclusive to the $1s$ state and likely extends to other energy states far from the chemical potential.
Finally, as discussed in the main text, we believe these inaccurate prediction would likely be fixed, at least in part, by including both self-consistency and vertex corrections to the $GW$ approximation~\cite{takada:2001}.

\subsection{Isochorically heated beryllium}
\label{sec:si:results:non_therm}

\begin{figure*}[h!]
  \includegraphics[width=\textwidth]{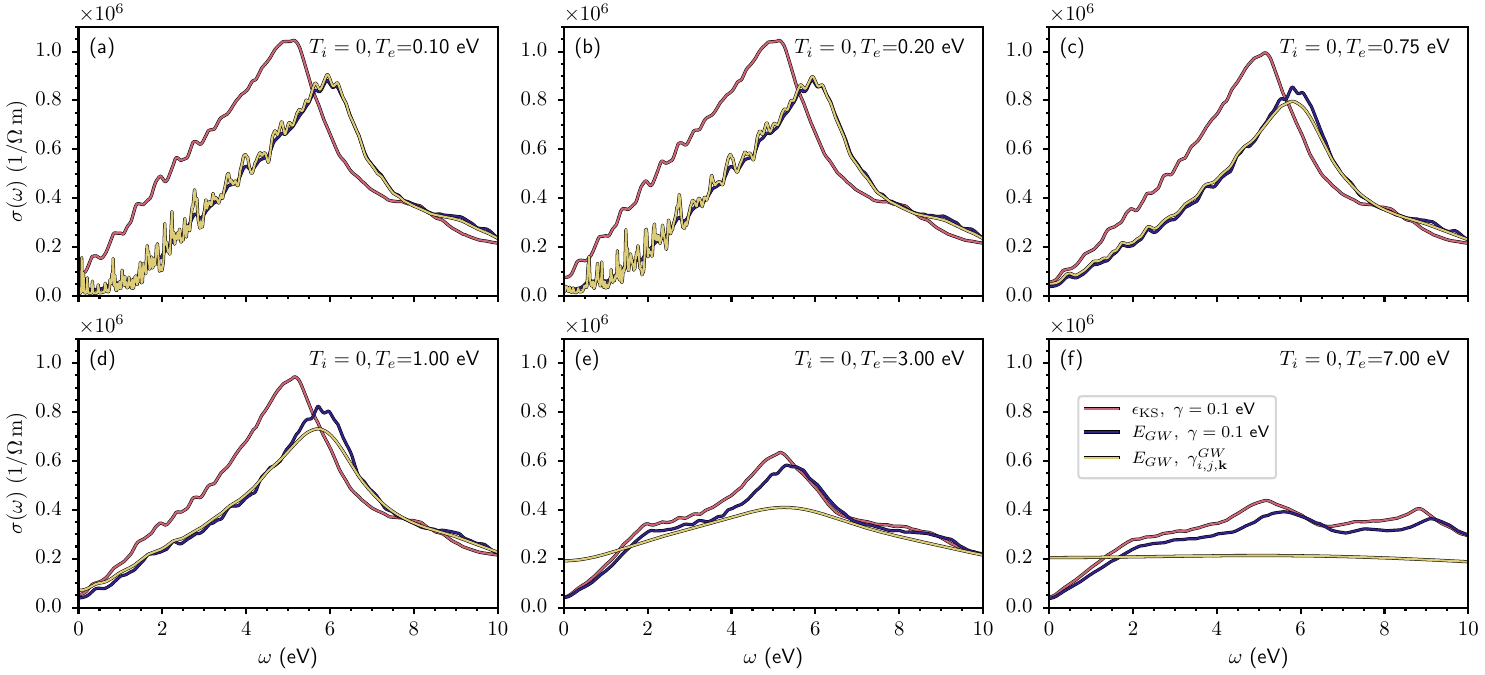}
  \caption{The conductivity traces predicted from standard KS-DFT (\ks), $GW$ corrections to the KS-DFT energies (\gw), and additionally $GW$-informed broadening (\gwD) for the non-thermalized ($T_i=0$) case, where the electronic temperatures are between $0.1$ and \SI{7}{\electronvolt}.
  }\label{fig:non_therm_sigma}
\end{figure*}

The isochorically heated case, where the ions occupy crystalline lattice sites and only the electronic temperature is elevated, requires fewer atoms and thus enables deeper analysis than the thermalized case considered in the main text.
Beyond this computational benefit, recent laser-heated experiments probe DC conductivities far from equilibrium~\cite{ofori-okai:2025}, where the electrons heat more rapidly than ions; thus,
first-principles DC conductivities for non-thermalized conditions can be useful.
We focus on the optical conductivities (see Fig.~\ref{fig:non_therm_sigma}) for electronic temperatures ranging from relatively cold conditions at $T_e=$~\SI{0.1}{\electronvolt} to warm dense conditions at $T_e=$~\SI{7}{\electronvolt}.
We pay special attention to two frequency regions: the zero-frequency limit and the frequencies near \SI{5}{\electronvolt}.
The zero-frequency limit is especially important to high-energy density science since the DC conductivity is a key parameter in hydrodynamic simulations~\cite{stanek:2024,haines:2024}.
However, as discussed in Sec.~\ref{sec:si:convergence},  $\mathbf{k}$-point sampling makes it very difficult to converge DC conductivities for a perfect crystal lattice (see Fig.~\ref{fig:nontherm_dc_kpoint} and Table~\ref{tab:nkpts_convergence}).
Thus, we only consider the qualitative trends of the DC conductivity in this section.
The other frequency region, near \SI{5}{\electronvolt}, exhibits a distinct non-Drude feature in the dynamic conductivity~\cite{li:2015} that is well-converged within KS-DFT but may pose theoretical challenges for atomic models.

\begin{figure}
  \includegraphics[width=\columnwidth]{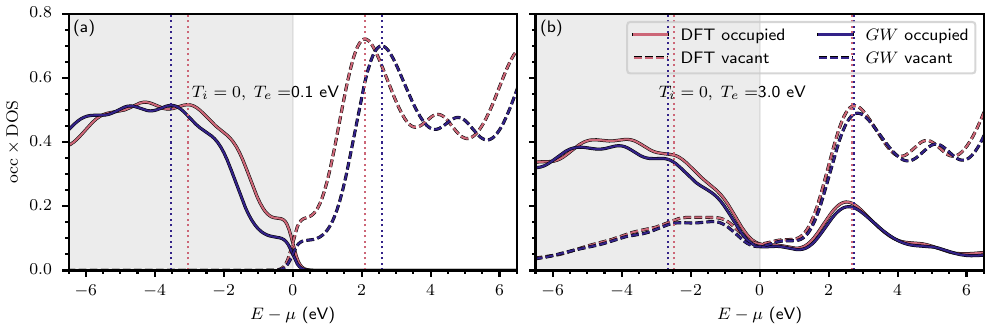}
  \caption{KS-DFT (red) and $GW$ (blue) densities of states weighted by Fermi occupation (solid) and vacancy (dashed) factors for \SI{1.97}{\gram\per\cubic\centi\meter} beryllium isochorically heated to electronic temperatures of (a) $0.1$ and (b) \SI{3}{\electronvolt}.
  The vertical lines correspond to the KS-DFT (solid) and $GW$ (dotted) estimated bounds of the pseudogap.
  }\label{fig:non_therm_dos}
\end{figure}

Similar to the main text, we begin by comparing KS-DFT and $GW$ predictions for the density of states (DOS) in Fig.~\ref{fig:non_therm_dos}.
While the DOS does not directly enter the KG formalism for conductivities, it provides insight into conductivity trends through the Onsager coefficients' dependence on eigenenergies and occupations (see Eq.~\eqref{eq:onsager_SI}).
Fig.~\ref{fig:non_therm_dos} shows exemplary results at $T_e=0.1$ and \SI{3}{\electronvolt}, where the DOS has been weighted by 
Fermi occupation and vacancy factors to give available populations for electronic transitions.
In isochorically heated conditions, the pseudogap near the Fermi energy is even more prominent and persists at higher temperatures than in the thermalized case of Fig.~\ref{fig:therm_dos_broad} in the main text.

Fig.~\ref{fig:non_therm_sigma} shows the effects of replacing the KS-DFT eigenvalues and occupations with QP eigenvalues and occupations and replacing the constant Lorentzian broadening parameter with a transition-dependent value.
Looking at these two regions, along with the overall conductivity behavior, we find that the influence of QP energies and transition-dependent broadening depends on the material conditions.
The effects of replacing KS-DFT eigenvalues and corresponding occupations with the $GW$ calculated QP eigenvalues and occupations, while keeping a constant Lorentzian broadening of \SI{0.1}{\electronvolt}, is most significant for temperatures below \SI{3}{\electronvolt} as illustrated in Fig~\ref{fig:non_therm_sigma}.
While the impact on the zero-frequency limit is small, we find in Fig.~\ref{fig:non_therm_sigma}(a-d) that for $0.1 \le T_e \le$~\SI{1}{\electronvolt}, KS-DFT and both $GW$ methods produce prominent peaks around \SI{5}{\electronvolt}.
For this temperature range, we find that the $GW$ peaks occur $0.56$\,--\,\SI{0.80}{\electronvolt} higher in energy than the KS-DFT peak, and these peak energy differences decrease as temperature increases, while for temperatures at $T_e=$~\SI{3}{\electronvolt} and above, we find that the KS-DFT and $GW$ curves with \SI{0.1}{\electronvolt} broadening have non-Drude peaks which are located at similar frequencies.
Replacing the KS-DFT values with $GW$ has a larger impact at lower temperatures not only for the non-Drude feature, but also all frequency values as demonstrated by the two curves tending towards agreement as the temperature increases in Fig.~\ref{fig:non_therm_sigma}.

\begin{figure}
  \includegraphics[width=\columnwidth]{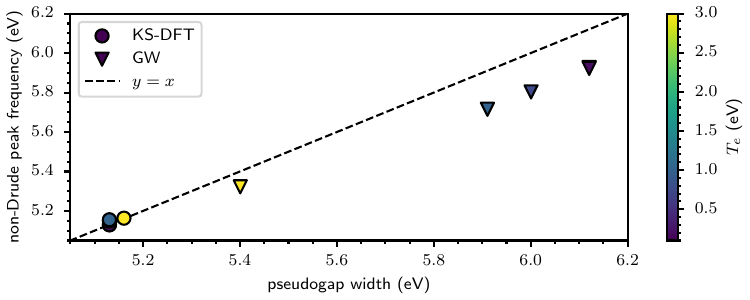}
  \caption{Non-Drude peak frequency as a function of estimated pseudogap width for KS-DFT (circles) and GW (triangles) calculations. 
  Point color indicates the assigned electronic temperature $T_e$, as shown by the colorbar.
  The dashed black line, $y=x$, serves to illustrate the near one-to-one relation between the peak frequency and pseudogap width.}
  \label{fig:pseudo_gap}
\end{figure}

The weighted DOS shown in Fig.~\ref{fig:non_therm_dos} explains the non-Drude behavior of the dynamic conductivities near $\omega=$~\SI{5}{\electronvolt}.
For both the low temperature case at $T_e=$~\SI{0.1}{\electronvolt} and the high temperature case at $T_e=$~\SI{3}{\electronvolt}, the expected beryllium pseudogap is clearly seen near the chemical potential.
The pseudogap is a key contribution to the non-Drude peak, where we find the difference between the KS-DFT and $GW$ pseudogap is directly associated with the non-Drude peak frequency differences.
Since the pseudogap is nontrivial to define, we estimate it by finding the local maxima of the DOS around the chemical potential.

The conductivity at $T_e=$~\SI{0.1}{\electronvolt} has a KS-DFT non-Drude peak that is approximately \SI{0.8}{\electronvolt} lower in frequency relative to the $GW$ feature.
In Fig.~\ref{fig:non_therm_dos}(a), we see the bounds of the pseudogap are estimated to be $-3.5$ below and \SI{2.6}{\electronvolt} above the chemical potential, giving a pseudogap width of about \SI{6.1}{\electronvolt}.
The $GW$ pseudogap is about \SI{1.0}{\electronvolt} larger than the KS-DFT prediction of \SI{5.1}{\electronvolt}, where the bounds are estimated to be $-3.0$ and \SI{2.1}{\electronvolt}.
Raising the electronic temperature to \SI{3}{\electronvolt}, we see that the KS-DFT and $GW$ pseudogap discrepancies diminishes.
The bounds of the $GW$ and KS-DFT gap are $-2.5$ to \SI{2.7}{\electronvolt} and $-2.7$ to \SI{2.7}{\electronvolt} respectively, reducing the pseudogap difference to \SI{0.24}{\electronvolt}.
The relationship between the non-Drude peak frequency and pseudogap width are illustrated Fig.~\ref{fig:pseudo_gap}, and we see the near-linear relation between the non-Drude peak frequency and pseudogap.
We find the KS-DFT pseudogap width to be both less temperature dependent and closer to the peak frequency location since the largest difference between the values is \SI{0.02}{\electronvolt} at \mbox{$T=\SI{1.00}{\electronvolt}$}.
Alternatively, the $GW$ results depend more on temperature, as well as the peak frequency and pseudogap having a less exact relation; where, the \mbox{$T=\SI{0.20}{\electronvolt}$} pseudogap is \SI{0.2}{\electronvolt} below the peak location.
Fig.~\ref{fig:pseudo_gap} also clearly illustrates the $GW$ \textit{opening} of the pseudogap.

\begin{figure}
  \includegraphics[width=0.5\columnwidth]{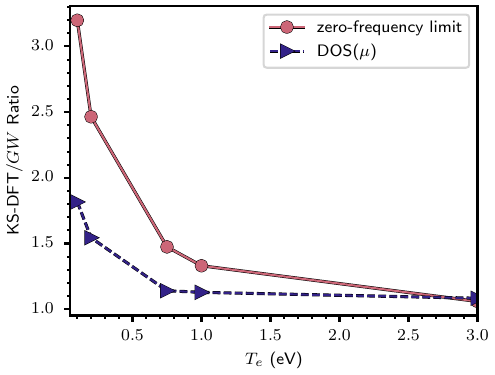}
  \caption{The ratio between KS-DFT and $GW$ predictions for the zero-frequency limit of the optical conductivity (red) and the DOS at the chemical potential (blue) as a function of electronic temperature for non-thermalized \SI{1.97}{\g\per\cubic\centi\meter} beryllium.
}\label{fig:non_therm_ratios}
\end{figure}

The qualitative trends of the optical conductivity in the zero-frequency (DC) limit is understood by looking at the number of thermally populated or vacant states available around the chemical potential.
At $T_e=$ \SI{0.1}{\electronvolt}, Fig.~\ref{fig:non_therm_dos}(a) shows that only states slightly above (below) the chemical potential are occupied (vacant) due to thermal excitations.
The number of thermally excited KS-DFT states at the chemical potential is about $1.5\times$ more than $GW$, while the KS-DFT optical conductivity is about $3\times$ as large as the $GW$ optical conductivity in the zero-frequency limit.
At $T_e=$ \SI{3}{\electronvolt}, the ratio of thermally occupied KS-DFT to $GW$ is about $1$, and we find a similar result for the KS-DFT and $GW$ zero-frequency limit ratios.
This trend is shown for all temperatures in Fig.~\ref{fig:non_therm_ratios}, where both the zero-frequency and DOS at mu ratios approach $1$ with increasing temperature.

\begin{figure}
  \includegraphics[width=0.80\columnwidth]{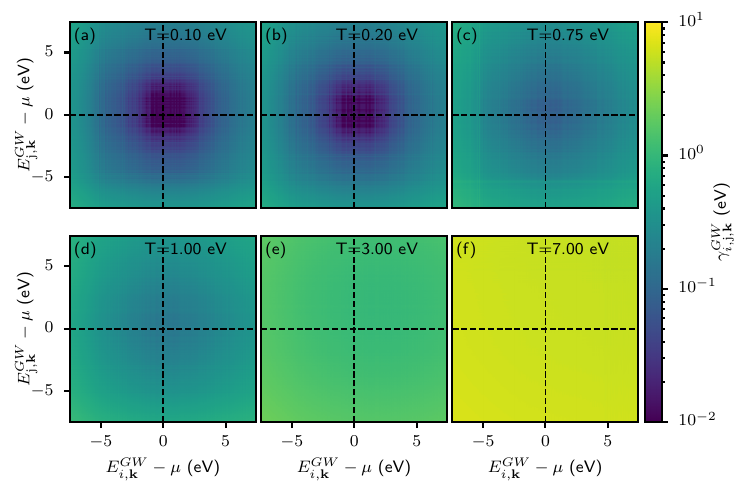}
    \caption{The electronic temperature-dependent $GW$-informed Lorentzian broadening of non-thermalized \SI{1.97}{\g\per\cubic\centi\meter} beryllium as function of the initial ($E_{i,\mathbf{k}}$) and final ($E_{j,\mathbf{k}}$) optical transition energy.
    }
\label{fig:non_therm_broad}
\end{figure}

We have found that QP energies and occupations have a greater effect at lower temperatures, while are almost negligible at high temperatures, however, the $GW$ approximation cannot be discarded at high temperature due to the significant impact that the $GW$-infromed broadening effects have on the conductivity.
Replacing the constant Lorentzian broadening value of \SI{0.1}{\electronvolt} in the KG equation with the physically informed $GW$ broadening, as defined by Eq.~\ref{eq:gamma_gw} and illustrated in Fig.~\ref{fig:non_therm_broad}, significantly changes the conductivity at high electronic temperatures.
The effects of the $GW$-infromed broadening parameter are elucidated when comparing the $GW$ curves in Fig.~\ref{fig:non_therm_sigma}.
Comparing these curves, we see that at electron temperatures between $0.1$ and \SI{1}{\electronvolt}, the two curves are in overall good qualitative agreement, where at $T_e=0.1$ and \SI{0.2}{\electronvolt} the $GW$ conductivities with $GW$ broadening are noisier than the $GW$ conductivities that only correct the KS-DFT energies.
This result can be explained by the fact that the $GW$-informed broadening values at this temperature tends to be about an order of magnitude lower than \SI{0.1}{\electronvolt} for energy states a few eV around the chemical potential.
Elevating the temperature to $T_e=0.75$ and \SI{1}{\electronvolt}, we start to see fully $GW$ informed conductivity predict higher conductivities in the zero-frequency limit, due to the $GW$ broadening values for thermally accessible energy states approaching and going beyond \SI{0.1}{\electronvolt}.
In Fig.~\ref{fig:non_therm_sigma}(e) and~\ref{fig:non_therm_sigma}(f), we see the significant effects of the $GW$-dependent broadening on the conductivity predictions at high electronic temperatures.
Compared to the KS-DFT or $GW$ curve with \SI{0.1}{\electronvolt} broadening, the $GW$ dynamic conductivity with $GW$ broadening is significantly broadened, so much so that at $T_e=$~\SI{7}{\electronvolt} it is nearly a constant value up to $\omega=$~\SI{20}{\electronvolt}.
We observe that the non-Drude feature at $T_e=$~\SI{3}{\electronvolt} is almost broadened out, while at $T_e=$~\SI{7}{\electronvolt} it is completely broadened out, where this large broadening of the non-Drude feature increases the conductivities in the zero-frequency limit over the KS-DFT or $GW$ values.
This behavior results from the $GW$ broadening exceeding \SI{0.1}{\electronvolt} by about one and two orders of magnitude for $T_e=$~\SI{3}{\electronvolt} and $T_e=$~\SI{7}{\electronvolt} respectively illustrated in Fig.~\ref{fig:non_therm_broad}(e) and (f).
Thus, the $GW$-informed Lorentzian broadening values governed by e-e scattering play a significant role at high electronic temperatures, however, we must focus on the low-temperature regime to better understand how the QP energies and occupations change the conductivity.

\end{document}